\documentclass[11pt]{article}
\usepackage{epsfig}
\newcommand{\sss}{\vspace{.2in}}

\newcommand{\be}{\begin{equation}}
\newcommand{\ee}{\end{equation}}
\newcommand{\bea}{\begin{eqnarray}}
\newcommand{\eea}{\end{eqnarray}}
\newcommand{\sn}{{\rm sn}}
\newcommand{\cn}{{\rm cn}}
\newcommand{\dn}{{\rm dn}}
\newcommand{\sech}{{\rm sech}}

\begin{document}
\vspace{.2in}
~\hfill{\footnotesize UICHEP-TH/01-8,~LA-UR-02-1064}
\vspace{.5in}
\begin{center}
{\LARGE {\bf Periodic Solutions of Nonlinear Equations
Obtained by Linear Superposition}}
\end{center}
\vspace{.5in}
\begin{center}
{\Large{\bf  \mbox{Fred Cooper}
 }}\\
\noindent
{\large Theoretical Division, Los Alamos National Laboratory, Los Alamos, NM 87545}
\end{center}
\vspace{.2in}
\begin{center}
{\Large{\bf  \mbox{Avinash Khare}\footnote{Permanent address: Institute of Physics, Sachivalaya Marg, Bhubaneswar 751005, Orissa, India}  and
   \mbox{Uday Sukhatme} 
 }}\\
\noindent
{\large Department of Physics, University of Illinois at Chicago, Chicago, IL 60607-7059}\\
\end{center}  
\sss
\vspace{1.2in}
\begin{center}
{\Large {\bf Abstract}}
\end{center}

\sss
We show that a type of linear superposition principle works for several nonlinear differential equations. Using this approach, we find periodic solutions of the Kadomtsev-Petviashvili (KP) equation, 
the nonlinear Schr{\" o}dinger (NLS) equation, the $\lambda \phi^4$ model, 
the sine-Gordon equation and the Boussinesq equation 
by making appropriate linear
superpositions of  known periodic solutions. This unusual procedure for 
generating solutions is successful as a consequence of some 
powerful, recently discovered,  cyclic identities satisfied by the Jacobi 
elliptic functions.
\newpage

\section{\bf Introduction}

The fact that Jacobi elliptic functions arise 
naturally as traveling wave solutions of many nonlinear systems has been known for quite some time 
(see for example \cite{dra}). Although for the solitary 
wave solutions of these nonlinear equations,
there is no superposition principle  
(except when the solitary waves are far apart),
for the periodic solutions
the situation turns out to be quite different. 
It has recently been 
shown \cite{kha6} that certain specific linear combinations of known 
periodic solutions of the
Korteweg-de Vries (KdV) and modified Korteweg-de Vries (mKdV) equations as well as $\lambda \phi^4$ theory, also satify these equations. This unexpected result is a 
consequence of some remarkable, recently established, identities involving 
Jacobi elliptic functions \cite{kha7}. Basically, the identities take the 
cross terms generated by the nonlinear terms in the differential equations and 
convert them into a manageable form. The purpose of this article is to show that 
such a procedure also works for other well-known nonlinear equations, namely 
the Kadomtsev-Petviashvili (KP) equation, the nonlinear Schr{\" o}dinger  
equation(NLSE), the sine-Gordon equation and the 
Boussinesq equation. It should be noted that the above list includes both integrable as well as nonintegrable systems.
These equations are of interest 
in several diverse areas of physics. The NLSE governs 
the propagation of an electromagnetic
wave in a glass fiber, or the spatial evolution of an electromagnetic 
field in a planar waveguide.
Temporal solitons described by the NLSE were first observed in 
1980 \cite{temp}, and the first confirmation and studies of 
spatial solitons in planar waveguides were reported  
in 1988 \cite{stat1, stat2}. 
Similarly, the $\lambda \phi^4$ and the sine-Gordon equations arise in several
condensed matter physics applications.
\sss
\section{\noindent \bf The Kadomtsev-Petviashvili (KP) Equation} 

The KP 
equation is a two-dimensional generalization of the KdV equation and is given by
\be\label{E1}
(u_t-6uu_x+u_{xxx})_x + 3u_{yy} = 0 ~.
\ee		
Properties of the KP equation are discussed in many texts \cite{dra}. In 
particular, the simplest, periodic, cnoidal traveling wave 
solution is
\be\label{E2}
u_1(x,y,t) = -2 \alpha^2 \dn^2(\xi_1,m)+\beta \alpha^2~,~~\xi_1 \equiv \alpha(x+\gamma \alpha y-b_1\alpha^2 t)~,
\ee
where $\alpha, \gamma, m$ and $\beta$ are constants, and the ``velocity" $b_1$ is given by
\be\label{E3}
b_1 = 8 - 4m -6\beta+3 \gamma^2 ~.
\ee			
In this article, for Jacobi elliptic functions, we use the standard notation 
$\dn \,(\xi,m),~\sn \,(\xi,m),~\cn \,(\xi,m)$, where $m$ is the elliptic 
modulus parameter $(0 \le m \le 1)$. The solution (2) remains unchanged 
when $x$ is increased by $2K(m)/\alpha$, where $K(m)$ is the complete 
elliptic integral of the first kind \cite{abr}. In the limiting case $m=1$ 
(and $\beta =0$), one recovers the familiar single soliton form
$-2 \alpha^2 \sech^2(\alpha(x+\gamma \alpha y-b_1\alpha^2 t))$.

We will make suitable linear combinations of solution (2) and show that the result is also a periodic solution of the KP equation. Our procedure consists of adding terms of the kind given in (2) but centered at $p$ equally spaced 
points along the period $2K(m)/\alpha$, where $p$ is any integer. The 
$p$-point solution is
\be\label{E4}
u_p(x,y,t) = -2 \alpha^2 \sum_{i=1}^{p} d_i^2 + \beta \alpha^2~; ~ 
d_i \equiv \dn[\xi_p + \frac{2(i-1)K(m)}{p},m]~, 
~\xi_p \equiv \alpha(x+\gamma \alpha y-b_p\alpha^2 t) ~.
\ee
Clearly, $p=1$ is the original solution, but for any other $p$, we have new 
expressions which, as we shall show, also solve the KP equation. For 
convenience, we define the quantities $s_i$ and $c_i$ in analogy to the 
quantity $d_i$ defined above:
\be\label{E5}
~ s_i \equiv \sn[\xi_p + \frac{2(i-1)K(m)}{p},m]~~,
~ c_i \equiv \cn[\xi_p + \frac{2(i-1)K(m)}{p},m]~~.
\ee 

\begin{figure}
   \centering
  \epsfig{figure=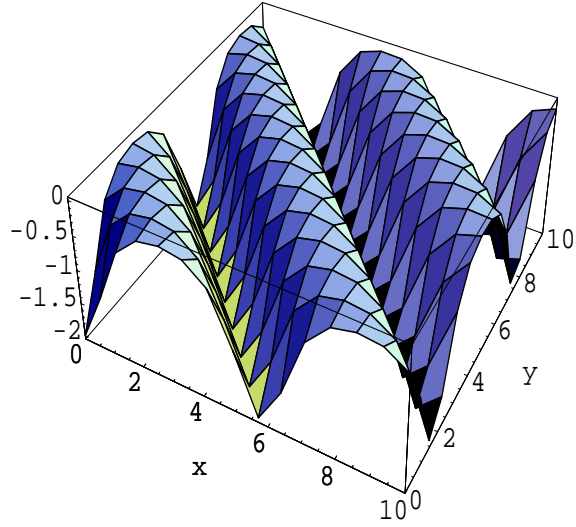,width=3 in,height=3 in}
\caption{$u_1(x,y,t=0)$ vs. $(x,y)$ for $m=0.95$, $\alpha=\gamma=1$, $\beta=0.$}
\label{fig:kpu1} 
\end{figure}

The KP equation contains the KdV operator $u_t-6uu_x+u_{xxx}$. It has been shown in detail in ref. \cite{kha6} that eq. (\ref{E4}) with $\gamma=0$ is a solution of the KdV equation. The proof is based on the identity
\be\label{E6}
\sum_{i<j}^{p} d_i^2 d_j^2 = 
A_1 (p,m) \sum_{i=1}^{p} d_i^2 + A_2 (p,m)~.
\ee
This is one of many
powerful new identities \cite{kha7} which reduce by 2 (or a larger even number)
 the degree of cyclic homogeneous polynomials in Jacobi elliptic functions. 
The constants $A_1(p,m)$ and $A_2(p,m)$ in identity (\ref{E6}) can be evaluated in general by 
choosing any specific convenient value of the argument $\xi$ of the Jacobi 
elliptic functions. The results for $A_1(p,m)$ for small values of $p$ are:
\be\label{E7}
A_1 (p=2,m)=0~~,~~A_1 (p=3,m)=\frac{-2(m-1+q^2)}{1-q^2}~~,
~~A_1 (p=4,m)=-2\sqrt{1-m}~~,
\ee
where
\be\label{E8}
q \equiv \dn (2K(m)/3,m)~.
\ee
The limiting values at $m=0,1$ are also particularly simple:
\be\label{E8a}
A_1 (p,m=0) = -\frac{1}{3}(p-1)(p-2)~;~	A_1 (p,m=1) = 0~.
\ee
Taking expression (\ref{E4}) and using identity (\ref{E6}), the left side of the KP equation 
(\ref{E1}) 
becomes
\be\label{E9}
4m\alpha^5\{8-4m-6\beta - b_p+12A_1 (p,m)\}\frac{d}{dx}\sum_{i=1}^p s_ic_id_i 
+12m\gamma \alpha^4 \frac{d}{dy}\sum_{i=1}^p s_ic_id_i~. 
\ee
Clearly, this vanishes if
the velocity is given by 
\be\label{E15}
b_p = 8 - 4m-6\beta +12 A_1 (p,m)+3\gamma^2 ~.
\ee
Thus for this choice of velocity, the KP equation is solved by our $p$-point 
expression (\ref{E4}). Effectively,  our solutions of the KP equation and 
the corresponding solutions of the KdV equation have a difference of 
$3\gamma^2$ in their velocities $b_p$. Note that as in the KdV case, 
the results for $b_p$ can be 
positive or negative depending on the values of the 
parameters \cite{kha6}.
The behavior of $u_1$ for the parameters $m=0.95$, $\alpha=\gamma=1$, $\beta=0$
is shown in fig. \ref{fig:kpu1}.

In addition to the solution (\ref{E2}), another well known periodic solution
of the KP eq. (\ref{E1}) of period $4K(m)$ is
\be\label{E16}
v_1(x,y,t) =  \alpha^2 \bigg [m \sn^2(\eta_1,m)
\pm \sqrt{m}\cn(\eta_1,m) \dn(\eta_1,m) \bigg ]~,
~~\eta_1 \equiv \alpha(x+\gamma \alpha y-q_1 \alpha^2 t)
\ee
with velocity $q_1 = (-1-m+3\gamma^2)$.

\begin{figure}
   \centering
  \epsfig{figure=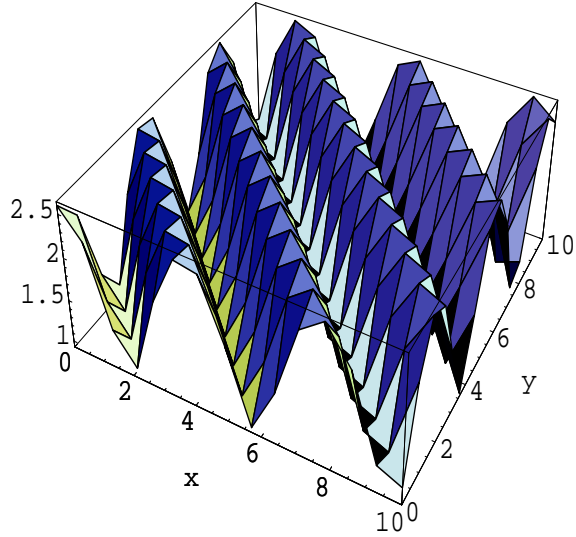,width=3. in,height=3. in}
\caption{$v_3(x,y,t=0)$ vs. $(x,y)$ for  $m=0.95$, $\alpha=\gamma=1.$}
\label{fig:v3}
\end{figure}

Starting from this solution, we can again obtain many additional periodic
solutions of the KP equation of period $4K(m)/p$ in case $p$ is an odd integer. The general $p$-point
solution is given by
\be\label{E17}
v_p(x,y,t)=\alpha^2 \sum_{i=1}^p ~[m{\tilde s}_i^2 
\pm \sqrt{m}\, {\tilde c}_i {\tilde d}_i]~,~~p ~~{\rm odd}~,
\ee
where we define
\be\label{E18} 
~ {\tilde s}_i \equiv \sn[\eta_p + \frac{4(i-1)K(m)}{p},m]~~,
~ {\tilde c}_i \equiv \cn[\eta_p + \frac{4(i-1)K(m)}{p},m]~~,
~ {\tilde d}_i \equiv \dn[\eta_p + \frac{4(i-1)K(m)}{p},m]~~.
\ee
As has been shown in detail in ref. \cite{kha6}, eq. (\ref{E17}) with $\gamma =0$
is a solution of the KdV equation with velocity 
\be\label{E19}
q_p = -(1+m)-6[B_1 (p,m)-C_1 (p,m)]~,
\ee
where the quantities $B_1 (p,m)$ and $C_1 (p,m)$ come from the following identities:
\be\label{E20}
m\sum_{i<j}^p {\tilde s}_i {\tilde s}_j = B_1 (p,m)~,~~
m\sum_{i<j<k}^p {\tilde s}_i {\tilde s}_j {\tilde s}_k = C_1(p,m)~\sum_{i=1}^p {\tilde s}_i ~.
\ee
It is easily checked that even in the KP case ($\gamma \ne 0$), eq. (\ref{E17}) is an exact
solution, 
the only difference being that the velocity in the KP case is larger by
$3\gamma^2$. As an illustration, for the $p=3$ case, it is easily shown that 
\cite{kha7}
\be\label{E21}
B_1 (3,m) = -(1-q^2)~,~~C_1 (3,m) = -m/(1-q^2)~,
\ee
so that the velocity of the KP soliton is given by
\be\label{E22}
q_3 = -1-m+6(1-q^2)-\frac{6m}{1-q^2} +3\gamma^2~,
\ee 
where $q$ has been defined in eq. (\ref{E8}). 
Note that for $p$ even, we do not obtain any new solutions. To illustrate our results, in fig. \ref{fig:v3} we plot $v_3(x,y)$ at time $t=0$ for the choice $\alpha = \gamma =1$ and $m=0.95$.  

\sss
\section{\noindent \bf The Nonlinear Schr{\" o}dinger Equation}

The NLSE with both attractive and repulsive nonlinearity 
has found many physical applications in several diverse areas
including fiber optics, Bose-Einstein condensates and waveguides \cite{5}. 

\subsection{\bf Case I: Attractive Nonlinearity}

The NLSE with attractive nonlinearity is given by ($\hbar = 2m =1$)
\be\label{E23}
iu_t + u_{xx} +u\mid u \mid^2 =0~,
\ee
where without any loss of generality we have fixed the coefficient of 
the nonlinear term to be unity. As usual, one starts with the ansatz \cite{dra}
\be\label{E24}
u(x,t) = r(\xi)e^{i(\theta (\xi)+nt)}~,~~\xi \equiv x-vt~,
\ee 
which on substituting in eq. (\ref{E23}) yields 
\be\label{E25}
\theta'(\xi) = \frac{1}{2}(v+\frac{A}{r^2})~,
\ee
\be\label{E26}
r'^2 (\xi) = -\frac{r^4}{2}+(n-\frac{v^2}{4})r^2 
-\frac{B}{2}-\frac{A^2}{4r^2}~,
\ee
where prime denotes a derivative with respect to the argument $\xi$ and 
$A$, $B$ are constants of integration. Thus the
whole problem reduces to finding the solutions of eq. (\ref{E26}), after which
$\theta$ is easily obtained by using eq. (\ref{E25}) and performing one integration. 

The well known soliton solution of eq. (\ref{E26}) is
\be\label{E27}
r(\xi) = \sqrt{2}~ \sech~\xi~,~~\theta = \frac{v\xi}{2}~, ~~A = B =0~, 
~v^2 = 4(n-1)~,
\ee
which is valid only for $n \ge 1$. It may be noted that
a somewhat more general solution with arbitrary amplitude $\alpha$ is
easily obtained, since if $u(x,t)$ is a solution of the NLSE,
then $\alpha u(\alpha x,\alpha^2 t)$ is also a solution of the same equation.

The two simplest, periodic, cnoidal traveling wave solutions of  
eq. (\ref{E26}) are $(\xi_1 = x-v_1 t)$
\be\label{E28}
r_1(\xi) = \sqrt{2} \, \dn~\xi_1~,~~\theta_1 = \frac{v_1 \xi_1}{2}~, 
~A =0~,~B = 4(1-m)~, 
~v_1^2 = 4(n+m-2)~,
\ee 
\be\label{E29}
r_1(\xi) = \sqrt{2m} \, \cn~\xi_1~,~~\theta_1 = \frac{v_1\xi_1}{2}~, 
~A =0~,~B = -4m(1-m)~,
~v_1^2 = 4(n+1-2m)~.
\ee 
In the limiting case $m=1$, one recovers the familiar soliton solution 
(\ref{E27}).

We shall now show that suitable linear combinations of the solutions 
(\ref{E28}) and (\ref{E29}) are also solutions of eq. (\ref{E26}). Consider
first the solution (\ref{E28}). Our
solutions consist of adding terms of the kind given 
in this equation but centered at $p$ equally spaced points along the period 
$2K(m)$, where $p$ is any integer. The $p$-point solution is
\be\label{E30}
r_p(x,t) = \sqrt{2} \sum_{i=1}^{p} d_i~,  ~ d_i \equiv \dn[\xi_p 
+ \frac{2(i-1)K(m)}{p},m]~, ~\xi_p \equiv (x-v_p t) ~.
\ee
Clearly, $p=1$ is the original solution, but for any other $p$, we have
solutions of period $2K(m)/p$. 

In order to verify that expression (\ref{E30}) is indeed a solution of eq. (\ref{E26}),
one needs the identities
\bea\label{E31}
&&(\sum_{i=1}^p d_i)^2 = \sum_{i=1}^p d_i^2 + A(p,m)~, \nonumber \\
&&(\sum_{i=1}^p d_i)^4 = \sum_{i=1}^p d_i^4 + C(p,m)\sum_{i=1}^p d_i^2 
+D(p,m)~, \nonumber \\
&&m^2\sum_{i<j}^p s_i c_i s_j c_j = E(p,m)\sum_{i=1}^p d_i^2 + F(p,m)~,
\eea
which can be easily established by following the procedure discussed 
in ref. \cite{kha7}. The general expression for the 
velocity $v_p$ is
\be\label{E32}
v^2_p = 4 \big [n+m-2-C(p,m)-2E(p,m) \big ]~.
\ee
Some explicitly computed values of the constants $C(p,m)$ and $E(p,m)$ are
\bea\label{E33}
&&C(2,m) = 4E(2,m) = 4\sqrt{1-m}~,~~C(3,m)= 4E(3,m)=\frac{8mq}{1-q^2}~, \nonumber\\
&&C(4,m) = 4E(4,m) = 4{\tilde t}(2+{\tilde t}+2{\tilde t}^2)~,
\eea
where $q$ is given by eq. (\ref{E8}) and ${\tilde t}$ is given by
\be\label{E33a}
{\tilde t} \equiv (1-m)^{1/4}~.
\ee 
On the other hand, for any $p$ at $m=0$, 
$C(p,0) =4E(p,0) = \frac{4(p^2-1)}{3}$ and at
$m=1$, $C(p,1) = E(p,1) = 0$. It then follows from eq. (\ref{E32})
that the solution
$r_p$ as given by eq. (\ref{E30}) is valid only if $n \ge 2p^2$, and in
this case $v^2$ changes from $4(n-2p^2)$ to $4(n-1)$ as $m$ goes from 0
to 1.  

For odd $p$, using the solution (\ref{E29}), we obtain the following solution
of the NLSE (\ref{E26}) by linear superposition
\be\label{E34}
r_p(x,t) = \sqrt{2m} \sum_{i=1}^{p} {\tilde c}_i~,  ~ 
{\tilde c}_i \equiv \cn[\eta_p 
+ \frac{4(i-1)K(m)}{p},m]~, ~\eta_p \equiv (x-v_p t) ~.
\ee
In order to verify that (\ref{E34}) is indeed a solution to the NLSE eq. (\ref{E26})
one needs the identities
\bea\label{E35}
&&(\sum_{i=1}^p {\tilde c}_i)^2 = \sum_{i=1}^p {\tilde c}_i^2 + G(p,m)~, \nonumber \\
&&(\sum_{i=1}^p {\tilde c}_i)^4 = \sum_{i=1}^p {\tilde c}_i^4 
+ H(p,m)\sum_{i=1}^p {\tilde c}_i^2 
+I(p,m) ~,\nonumber \\
&&m^2\sum_{i<j}^p {\tilde s}_i {\tilde d}_i {\tilde s}_j {\tilde d}_j 
= J(p,m)\sum_{i=1}^p {\tilde c}_i^2 + K(p,m)~,
\eea
which can be established following the procedure discussed in ref. \cite{kha7}. 
The general expression for the
velocity $v_p$ is
\be\label{E36}
v^2_p = 4 \big [n+1-2m-mH(p,m)-2J(p,m) \big ]~.
\ee
Some explicitly computed values of the constants $H(p,m)$ and $J(p,m)$ are
\be\label{E37}
mH(3,m)= 4J(3,m) = -4q[\frac{q+2}{(1+q)^2} +\frac{q}{1-q^2}]~,
\ee
where $q$ is given by eq. (\ref{E8}). 
Thus $v^2$ varies from $4(n+9)$ to $4(n-1)$ as the elliptic modulus parameter $m$ changes from 0 to 1.

On the other hand, for even $p$, we have obtained the following solution
of the NLSE eq. (\ref{E26}): 
\be\label{E38}
r_p(x,t) = \sqrt{2} \sum_{i~{\rm odd}}^{p} [d_i - d_{i+1}]~.   
\ee
In order to verify that eq. (\ref{E38}) is a solution of the NLSE eq. (\ref{E26})
one needs the identities
\bea\label{E39}
&& \bigg (\sum_{i~{\rm odd}}^p [d_i - d_{i+1}] \bigg )^2 = 
\sum_{i=1}^p d_i^2 + P(p,m)~, \nonumber \\
&& \bigg (\sum_{i~{\rm odd}}^p [d_i -d_{i+1}] \bigg )^4 
= \sum_{i=1}^p d_i^4 
+ L(p,m)\sum_{i=1}^p d_i^2 
+M(p,m) ~,\nonumber \\
&&m^2 \bigg [\sum_{i+j~{\rm even}}^p s_i c_i s_j c_j 
-\sum_{i+j~{\rm odd}}^p s_i c_i s_j c_j \bigg ] 
= N(p,m)\sum_{i=1}^p d_i^2 + Q(p,m)~,
\eea
which can be established following the procedure discussed in ref. \cite{kha7}. The general expression for the velocity $v_p$ is
\be\label{E40}
v^2_p = 4 \bigg [n+m-2-L(p,m)-2N(p,m) \bigg ]~.
\ee
Some explicitly computed values of the constants $L(p,m)$ and $N(p,m)$ are
\be\label{E41}
L(2,m)= 4N(2,m) = -4\sqrt{1-m}~; 
~~L(4,m)= 4N(4,m) = -4{\tilde t}(2-{\tilde t}+2{\tilde t}^2)~, 
\ee
where ${\tilde t}$ is as given by eq. (\ref{E33a}).
Thus for $p =2~[4]$, $v^2$ varies from $4(n+4)~ [4(n+16)$] 
to $4(n-1)$ as $m$ changes from 0 to 1.
Generalizing the results in eqs. (\ref{E34}) or (\ref{E38}), one finds 
that $v^2$ varies
from $4(n+p^2)$ to $4(n-1)$ as $m$ changes from 0 to 1.

\subsection{\bf Case II: Repulsive Nonlinearity}

The NLSE with repulsive nonlinearity is given by 
\be\label{E42}
iu_t + u_{xx} -u\mid u \mid^2 =0~,
\ee
We again start with the ansatz given by eq. (\ref{E24}) and on following the
same steps as given in eqs. (\ref{E24}) to (\ref{E26}) it is easily seen that
the $\theta$ equation (eq. (\ref{E25})) is the same as before while the $r$ equation
is almost the same except for the sign of the $r^4$ term. In particular, the $r$
equation is now given by
\be\label{E43}
r'^2 (\xi) = \frac{r^4}{2}+(n-\frac{v^2}{4})r^2 
-\frac{B}{2}-\frac{A^2}{4r^2}~,
\ee

The well known soliton solution to this equation is
\be\label{E44}
r(\xi) = \sqrt{2} \, {\rm tanh}~\xi~,~~\theta = \frac{v\xi}{2}~,~~ A =0~,
~~B = -4~, 
~v^2 = 4(n+2)~.
\ee
The simplest, periodic, cnoidal traveling wave solution to
eq. (\ref{E43}) is 
\be\label{E45}
r_1(\xi) = \sqrt{2m} \, \sn~\xi_1~,~~\theta_1 = \frac{v_1 \xi_1}{2}~, 
~~A =0~,~B = -4m~, 
~v_1^2 = 4(n+1+m)~.
\ee 

For odd $p$, using the solution eq. (\ref{E45}), we obtain the following solutions
of the NLSE eq. (\ref{E26}) by linear superposition
\be\label{E46}
r_p(x,t) = \sqrt{2m} \sum_{i=1}^{p} {\tilde s}_i~,  ~ 
{\tilde s}_i \equiv \sn[\eta_p 
+ \frac{4(i-1)K(m)}{p},m]~, ~\eta_p \equiv (x-v_p t) ~.
\ee
In order to verify that eq. (\ref{E46}) is a solution to the NLSE eq. (\ref{E43})
one needs the identities
\bea\label{E47}
&&(\sum_{i=1}^p {\tilde s}_i)^2 = \sum_{i=1}^p {\tilde s}_i^2 + R(p,m)~, \nonumber \\
&&(\sum_{i=1}^p {\tilde s}_i)^4 = \sum_{i=1}^p {\tilde s}_i^4 
+ S(p,m)\sum_{i=1}^p {\tilde s}_i^2 
+T(p,m)~, \nonumber \\
&&\sum_{i<j}^p {\tilde c}_i {\tilde d}_i {\tilde c}_j {\tilde d}_j 
= U(p,m)\sum_{i=1}^p {\tilde s}_i^2 + Y(p,m)~,
\eea
which can be established by following \cite{kha7}. The general expression for the 
velocity $v_p$ is
\be\label{E48}
v^2_p = 4 \big [n+1+m+mS(p,m)-2U(p,m) \big ]~.
\ee
Some explicitly computed values of the constants $H(p,m)$ and $J(p,m)$ are
\be\label{E49}
mS(3,m)= -4U(3,m) = 4m[\frac{1}{1-q^2} -\frac{1-q^2}{m}]~,
\ee
where $q$ is given by eq. (\ref{E8}). 
Thus $v^2$ changes from $4(n+9)$ to $4(n+2)$ as $m$ changes from 0 to 1.

For even integer $p$, the linear superposition of elementary solutions does
not work. However, remarkably enough we find that the products of elementary 
solutions are also solutions. For example, the solution for $p=2$ is  
\be\label{E50}
r_2(x,t) = \sqrt{2}m s_1 s_2~,  ~ 
\ee
and the corresponding velocity is given by $v_2^2 = 4(n+4-2m)$. 
Generalization to higher even values of $p$ is straightforward. 

\subsection{\bf Solutions With $A \ne 0$}

It may be noted that since $A=0$
for all the solutions discussed so far, the expressions for
$\theta$ were rather trivial.
One way of obtaining a solution with
$A \ne 0$ is to start with the ansatz 
\be\label{E54}
r^2(\xi) = 2 \dn^2~\xi+\alpha~,
\ee
where $\alpha$ is a constant.   
It is easily checked that (\ref{E54}) is a solution to the NLSE 
(\ref{E26}) provided
\bea\label{E55}
&& \alpha = \frac{2}{3}(n-2+m-\frac{c^2}{4})~, A^2 =4\alpha \big [\alpha^2 
-(n-\frac{c^2}{4})\alpha +2(1-m) \big ]~, \nonumber\\
&& B=-3\alpha^2 + 4(n-\frac{c^2}{4})\alpha -4(1-m)~.
\eea

Starting from the solution (\ref{E54}) we can obtain a class of solutions by 
an appropriate linear superposition. 
For example, the 2-point solution is
\be\label{E56}
r^2(\xi) = 2 (d_1^2 +d_2^2) +\alpha~,
\ee
It is easily checked that this is indeed a solution provided
\bea\label{E57}
&& \alpha = \frac{2}{3}(n-2+m-\frac{c^2}{4})~, 
~~B=-3\alpha^2 + 4(n-\frac{c^2}{4})\alpha -16(1-m)~, \nonumber\\
&&A^2 =4 \bigg [\alpha^3 
-(n-\frac{c^2}{4}\alpha^2 +8(1-m)\alpha+8(1-m)(n+2-m-\frac{c^2}{4}) \bigg ]~. 
\eea
Generalization to arbitrary $p$ is straightforward.

\sss
\section{\noindent \bf The $\lambda \phi^4$ Model}

The kink (domain wall) solutions to the $\lambda \phi^4$ field theory 
in $(1+1)$ 
dimensions 
\be\label{E58}
\phi_{xx} - \phi_{tt} = \lambda \phi (\phi^2 -a^2)~,
\ee
have been widely discussed in the literature \cite{raj}. 
The famous static kink solution
is 
\be\label{E59}
\phi(x) = a \tanh (\sqrt{\lambda /2} \, ax)~,
\ee
from which the time-dependent solution 
\be\label{E60}
\phi(x,t) = a \tanh \bigg [\sqrt{\frac{\lambda}{2(1-v^2)}} \, a(x-vt) \bigg]~,
\ee
is immediately obtained by Lorentz boosting. Therefore, to begin with,
we shall   
discuss only the static periodic kink solutions in case $v^2 < 1$ 
(and $\lambda > 1$). 
Later, we shall discuss
time dependent solutions with $v^2 > 1$ (or $\lambda < 1$). 

\subsection{Static Periodic Kink Solutions}

It is well known that the static periodic kink solution to the field 
eq. (\ref{E58}) is
\be\label{E61}
\phi_1 (x) = \sqrt{\frac{2m}{1+m}} a ~\sn (\eta_1,m)~,~\eta_1 \equiv 
\sqrt{\frac{\lambda}{1+m}} \, ax~.
\ee

For any odd integer $p$, we find the following static kink solutions of the
$\lambda \phi^4$ field theory by a specific linear superposition of the basic
solution eq. (\ref{E61}):
\be\label{E62}
\phi_p (x) = \sqrt{2m} \, \alpha a \sum_{i=1}^p {\tilde s}_i~, ~p ~~ {\rm odd}~,
\ee
where
${\tilde s}_i, {\tilde c}_i, {\tilde d}_i$ are as given in eq. (\ref{E18})
with $\eta_p \equiv \sqrt{\lambda} \, \alpha ax$.
In order to verify that eq. (\ref{E62}) is a static periodic kink 
solution to the  $\lambda \phi^4$ theory
field eq. (\ref{E58}),
one needs the identity \cite{kha7}
\be\label{E63}
(\sum_{i=1}^p {\tilde s}_i)^3 = \sum_{i=1}^p {\tilde s}_i^3 
+ V(p,m) \sum_{i=1}^p {\tilde s}_i~.
\ee
The constant $\alpha$ is given by
\be\label{E64}
\alpha = \frac{1}{\sqrt{1+m+2mV(p,m)}}~.
\ee
As an illustration, consider $p=3$. $V(3,m)$ is given by 
\be\label{E65}
mV(3,m) = 3\bigg [\frac{m}{1-q^2} - (1-q^2) \bigg ]~,   
\ee
where $q$ is defined in eq. (\ref{E8}). 
Note that $\alpha$ varies from 
$1/3$ to $1/\sqrt{2}$ as $m$ varies from 0 to 1.

\begin{figure}
   \centering
  \epsfig{figure=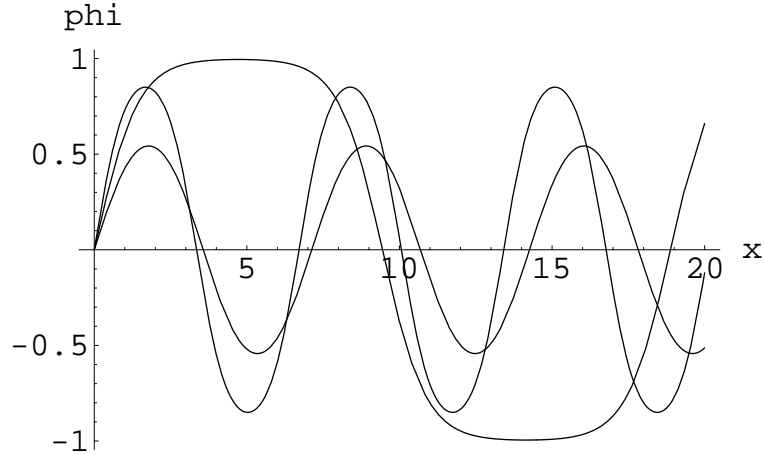,width=4 in,height=2.5 in}
\caption{$\phi_1(x)$ (largest amplitude curve), $\phi_2(x)$ and $\phi_3(x)$ (smallest amplitude curve)  for  $m=0.98$, $\lambda=a=1$.}
\label{fig:phi4}
\end{figure}

For even integer $p$, the linear superposition of elementary solutions does
not work. However, remarkably enough we find that even in this case, 
the products of elementary 
solutions are solutions. For example, the solution for $p=2$ is  
\be\label{E66}
\phi_2(x) = \sqrt{2} \, m \alpha a s_1 s_2~,  ~ 
\ee
where $s_{1,2}$ are as defined in eq. (\ref{E5}) with 
$\xi \equiv \sqrt{\lambda} \, \alpha ax$.
Using the identities derived in ref. \cite{kha7}, it is easily shown that (\ref{E66})
is a static periodic kink solution to the field eq. (\ref{E58}) provided
\be\label{E67}
\alpha = \frac{1}{\sqrt{2(2-m)}}~.
\ee
Note that $\alpha$ varies from $1/2$ to $1/\sqrt{2}$ as $m$ varies from 0 to 1.
Generalization to arbitrary even $p$ is straightforward.
It is thus clear that for arbitrary integer $p$, 
for static periodic kink solutions of 
$\lambda \phi^4$ theory, $\alpha$ will vary from $1/p$ to $1/\sqrt{2}$ as
$m$ varies from 0 to 1.
For the values $m=0.98$ and $\lambda=a=1$, we plot $\phi_1(x)$ from eq. (\ref{E61}), $\phi_2(x)$ from eq.
(\ref{E66}) and $\phi_3(x)$ from eq. (\ref{E62})  in
fig. \ref{fig:phi4}.
  
\subsection{Periodic Time-Dependent Kink Solutions}
 
While in the relativistic field theory context with $\lambda >0$, eqs. 
(\ref{E59}) and  (\ref{E61}) are the only solutions of $\lambda \phi^4$
field theory, in the condensed matter physics context, where velocity $v$ can exceed
velocity of sound (optical modes), or for relativistic case with 
$\lambda < 0$, one also has another soliton solution given by
\be\label{E68}
\phi (x,t) = \sqrt{2}a \, \sech(\beta (x-vt))~, 
~~~\beta = a \, \sqrt{\frac{\lambda}{(v^2-1)}}~, 
\ee
which is real if either $\lambda < 0$ (and $v^2 < 1) $ or $ v^2 >1$  (and 
$ \lambda > 0) $. 

The corresponding periodic soliton solutions to eq. (\ref{E58}) 
are well known and are 
given by \cite{aub}
\be\label{E69}
\phi_1 (\xi) = \sqrt{\frac{2}{2-m}} \, a \, \dn (\xi_1, m)~,
~\xi_1 =a\sqrt{\frac{\lambda}{(2-m)(v_1^2 -1)}}(x-vt)~,
\ee 
\be\label{E70}
\phi_1 (\eta) = \sqrt{\frac{2m}{2m-1}}a  \, \cn (\eta_1, m)~,
~\eta_1 =a\sqrt{\frac{\lambda}{(2m-1)(v_1^2 -1)}}(x-vt)~.
\ee
Notice that the solution (\ref{E70}) is valid only for $1/2 < m <1$ and both
solutions (\ref{E69}) and (\ref{E70}) are only valid for $v^2 >1$ 
(and $\lambda > 0$)  
or $\lambda < 0$ (and $v^2 < 1$).

Appropriate linear superposition of solutions (\ref{E69}) and
(\ref{E70}) are also periodic 
time-dependent kink solutions. For example, by using the linear superposition
of solutions (\ref{E69}),
we have the following solution to the field eq. (\ref{E58}), 
which is valid for any integer $p$:
\be\label{E71}
\phi_p(\xi) = \sqrt{2} \, a \alpha \sum_{i=1}^p d_i~,
\ee
where $d_i$ is as defined in eq. (\ref{E5}) with 
$\xi_p = a\sqrt{\frac{\lambda}{v_p^2 -1}} \alpha (x -v_p t)$. 
In order to prove that this is a solution, one needs the identity
\be\label{E72}
(\sum_{i=1}^p d_i)^3 = \sum_i^p d_i^3 + W(p,m) \sum_i d_i~,
\ee
which is easily proved following ref. \cite{kha7}. Using this identity one finds
that (\ref{E71}) is a solution provided
\be\label{E73}
\alpha^2 = \frac{1}{[2-m +W(p,m)]}~.
\ee
Some explicitly computed values of $W(p,m)$ are
\be\label{E74}
W(2,m)= 3\sqrt{1-m}~,~W(3,m)= \frac{6mq}{1-q^2}~,~W(4,m)= 3{\tilde t}[2+{\tilde t}+2{\tilde t}^2]~,
\ee
where ${\tilde t}$ is as defined in eq. (\ref{E33a}) and $q$ is given by eq. (\ref{E8}).
Further, $W(p,0) = p^2 -1$~, while $W(p,1) =0$.

For any odd integer $p$, we also have the following solution to the field 
eq. (\ref{E58}) by linear superposition
\be\label{E75}
\phi(\eta) = \sqrt{2m} \, a \alpha \sum_{i=1}^p {\tilde c}_i~,
\ee
where ${\tilde c}_i$ is as defined in eq. (\ref{E18}) with 
$\eta_p = a\sqrt{\frac{\lambda}{v_p^2 -1}} \alpha (x -v_p t)$. 
In order to prove that this is a solution, one needs 
the identity \cite{kha7}
\be\label{E76}
(\sum_{i=1}^p {\tilde c}_i)^3 = \sum_i^p {\tilde c}_i^3 
+ X(p,m) \sum_{i=1}^p {\tilde c}_i~.
\ee
Using this identity one finds
that (\ref{E75}) is a solution provided
\be\label{E77}
\alpha^2 = \frac{1}{[2m-1+2mX(p,m)]}~.
\ee
For example, one can check \cite{kha7} that 
\be\label{E77a}
X(3,m) = -6(1-m+q) +\frac{6q^2}{1-q^2}~, 
\ee
with $q$ being given by eq. (\ref{E8}), so that unlike the $p=1$ case, this
is an acceptable solution for all values of $m ~(0 \le m \le 1$). 

Similarly, for even integer $p$, we have solutions of the form
\be\label{E78}
\phi(\xi) = \sqrt{2} \, a \alpha \sum_{i~odd}^p [d_i -d_{i+1}]~,
\ee
where $d_i$ is as defined in eq. (\ref{E5}) with 
$\xi_p = a\sqrt{\frac{\lambda}{v_p^2 -1}} \alpha (x -v_p t)$. 
Unfortunately, it appears that 
all these solutions are only valid in a very narrow
range of values of $m$ corresponding to real values of $\alpha$. 

\section{\bf Sine-Gordon Field Theory}

In recent years, both sine-Gordon and $\lambda \phi^4$ field theory have
received considerable attention \cite{dra,raj}. 
In particular, sine-Gordon theory is the only 
relativistically invariant field theory having true soliton solutions. The equation under consideration is
\be\label{E80}
\phi_{xx} - \phi_{tt} = \sin \phi~.
\ee 
\subsection{\bf Static Soliton Solution by Linear Superposition}

The well known static one-soliton solution of this equation is given by
\be\label{E81}
\phi(x) = 4 \tan^{-1} e^{\pm x}~.
\ee
The corresponding time-dependent solution is easily obtained by Lorentz 
boosting and hence without any loss of generality we shall restrict our 
discussion to the static solution only (except when $v^2 > 1$). 
The solution (\ref{E81}) can also be written in the alternative form
\be\label{E82}
\sin (\frac{\phi(x)}{2}) = \sech~x~~.  
\ee
The two corresponding periodic static soliton solutions are well known and are
given by \cite{dra}
\be\label{p1sg}
\sin (\frac{\phi(x)}{2}) = \dn (x,m)~,  
\ee
\be\label{E84}
\sin (\frac{\phi(x)}{2}) = \cn (x/\sqrt{m},m)~, ~m > 0~.
\ee

For any odd integer $p$, we obtain the following periodic static soliton
solutions by linear superposition:
\be\label{E85}
\sin (\frac{\phi(x)}{2}) = \alpha \sum_{i=1}^p {\tilde d}_i~,  
\ee
where ${\tilde d}_i$ is as defined in eq. (\ref{E18}) with 
$\eta_p \equiv \alpha x$, while $\alpha$ is given by 
\be\label{E86}
\alpha^2 = \frac{1}{[p+A(p,m) + mR(p,m)]}~.
\ee
Here $A(p,m)$ and $R(p,m)$ are as defined by eq. (\ref{E31}) and (\ref{E47})
respectively, and with this choice of $\alpha$, one gets
\be\label{E86a}
\cos (\frac{\phi(x)}{2}) = \sqrt{m} \, \alpha \sum_{i=1}^{p} {\tilde s}_i~.
\ee
Note that use has been made of the identities \cite{kha7}
\be\label{E85a}
{\tilde s}_1 [{\tilde c}_2 +...+{\tilde c}_p] + c.p. = 0~,~
{\tilde s}_1 [{\tilde d}_2 +...+{\tilde d}_p] + c.p. = 0~,~
{\tilde d}_1 [{\tilde c}_2 +...+{\tilde c}_p] + c.p. = 0~,
\ee
in proving that (\ref{E85}) is indeed a solution to the field
eq. (\ref{E80}). 
For $p=3$ the values of the constants are
\be\label{E87}
A(3,m) = 2q(q+2)~,~mR(3,m) = 2(q^2-1)~,
\ee
where $q$ is given by eq. (\ref{E8}), 
so that $\alpha = 1/(1+2q)$ changes from $1/3$ to $1$ as $m$ varies
from 0 to 1. 

Another solution valid for any odd integer $p$ is
\be\label{E88}
\sin (\frac{\phi(x)}{2}) = \alpha \sum_{i=1}^p {\tilde c}_i~,  
\ee
where ${\tilde c}_i$ is as defined in eq. (\ref{E18}) with 
$\eta_p \equiv \alpha x/\sqrt{m}$. This solution is strictly valid only
if $m > 0$.
It is easily checked that this is indeed a solution to the field
eq. (\ref{E80}) provided
\be\label{E89}
\alpha^2 = \frac{1}{[p+G(p,m) + R(p,m)]}~,
\ee
where $G(p,m)$ and $R(p,m)$ are as defined by eq. (\ref{E27}) and (\ref{E43})
respectively. Note that with this choice of $\alpha$
\be\label{E89a}
\cos (\frac{\phi(x)}{2}) = \alpha \sum_{i=1}^{p} {\tilde s}_i~.
\ee
For $p=3$, $G(3,m)$ is given by 
\be\label{E90}
G(3,m) = -\frac{2q(q+2)}{(1+q)^2}~,
\ee
where $R(3,m)$ is as given by eq. (\ref{E87}), 
so that $\alpha = \frac{1+q}{1-q}$ ($m > 0$). 

\begin{figure}
   \centering
  \epsfig{figure=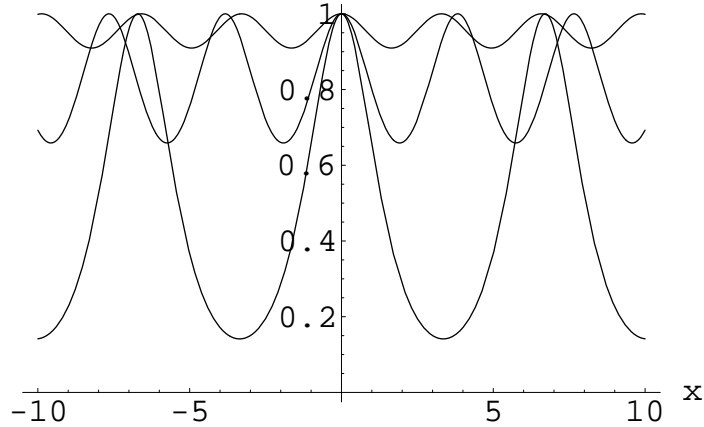,width=4.0 in,height=2.5 in}
\caption{$\sin[\phi(x)/2]$  for  $p=1,2,3$ . Increasing $p$ decreases the amplitude and
increases the frequency.}
\label{fig:sing2}
\end{figure}

Similarly, for $p =2$ we have the solution
\be\label{E91}
\sin (\frac{\phi(x)}{2}) = \alpha \sum_{i=1}^2 d_i~,  
\ee
where $d_i$ is as defined in eq. (\ref{E5}) with 
$\xi_2 \equiv \alpha x$. 
It is easily checked that this is indeed a solution of the field
eq. (\ref{E80}) provided
\be\label{E92}
\alpha = \frac{1}{1+\sqrt{1-m}}~,
\ee
so that $\alpha$ varies between 1/2 and 1 as $m$ changes from 0 to 1. 
Note that with this choice of $\alpha$
\be\label{E89b}
\cos (\frac{\phi(x)}{2}) = m \alpha s_1 s_2~.
\ee
In fig. \ref{fig:sing2}, we have plotted $\sin[\phi/2]$ vs. $x$  for $p=1,2,3$
corresponding to the right hand side of eqs. (\ref{p1sg}), (\ref{E91}) and (\ref{E85}).

Another solution for $p=2$ is 
\be\label{E93}
\sin (\frac{\phi(x)}{2}) = \alpha  [d_1 - d_2]~,  
\ee
where $d_i$ is as defined in eq. (\ref{E5}) with 
$\xi_2 \equiv \alpha x$. 
It is easily checked that this is indeed a solution to the field
eq. (\ref{E80}) provided $0 < m \le 1$ since $\alpha$ given by
\be\label{E94}
\alpha = \frac{1}{1-\sqrt{1-m}}~,
\ee
diverges at $m=0$. 

\subsection{\bf Periodic Time-Dependent Solutions}

As in the $\lambda \phi^4$ field theory case, in this case also we get
solutions by linear superposition, which are only valid 
for $v^2 > 1$. In particular, if $p$ is any odd integer, the 
periodic time-dependent solution is given by
\be\label{E94a}
\sin (\frac{\phi(x)}{2}) = \sqrt{m} \alpha \sum_{i=1}^p {\tilde s}_i~,  
\ee
where ${\tilde s}_i$ is as defined in eq. (\ref{E18}) with 
$\eta_p \equiv \alpha (x-vt) /\sqrt{v^2 -1}$. 
It is easily checked that this is indeed a solution of the field
eq. (\ref{E80}) provided $\alpha^2$ is again given by eq. (\ref{E86}).
Note that with this choice of $\alpha$ 
\be\label{E89c}
\cos (\frac{\phi(x)}{2}) = \alpha \sum_{i=1}^p {\tilde d}_i~.
\ee

Another solution for any odd integer $p$ is
\be\label{E95}
\sin (\frac{\phi(x)}{2}) = \alpha \sum_{i=1}^p {\tilde s}_i~,  
\ee
where ${\tilde s}_i$ is as defined in eq. (\ref{E18}) with 
$\eta_p \equiv \alpha (x-vt) /\sqrt{m(v^2 -1)}$. 
Thus, this solution is strictly valid only
if $m > 0$.
It is easily checked that this is indeed a solution of the field
eq. (\ref{E80}) provided $\alpha^2$ is as given by eq. (\ref{E89}).
Note that with this choice of $\alpha$,
\be\label{E89d}
\cos (\frac{\phi(x)}{2}) = \alpha \sum_{i=1}^p {\tilde c}_i~.
\ee

Finally, for $p =2$ we have the solution
\be\label{E96}
\sin (\frac{\phi(x)}{2}) = m \alpha s_1 s_2~,  
\ee
where $s_{1,2}$ is as defined in eq. (\ref{E5}) with 
$\xi_2 \equiv \alpha (x-vt)/\sqrt{v^2 -1}$. 
It is easily checked that this is indeed a solution of the field
eq. (\ref{E80}) provided $\alpha^2$ satisfies eq. (\ref{E91}).
Note that with this choice of $\alpha$
\be\label{E89e}
\cos (\frac{\phi(x)}{2}) = \alpha \sum_{i=1}^2 d_i~.
\ee

\section{\bf Boussinesq Equation}

The Boussinesq equation is given by
\be\label{E97}
u_{tt} - u_{xx} +3(u^2)_{xx} - u_{xxxx} = 0~.
\ee
The periodic one soliton solution of this equation is known to be
\be\label{E98}
u(x,t) = -2 \alpha^2 \dn^2 (\alpha (x-vt)) + \beta \alpha^2~,
\ee
where 
\be\label{E98a}
2(4-2m -3\beta)\alpha^2 = v^2 -1~. 
\ee
Thus $v^2 < 1$ if $\beta \ge 4/3$ and $v^2 > 1$ if $\beta \le 2/3$.
For $2/3 < \beta < 4/3, v^2$ changes sign at some value of $m~ (0 \le m \le 1$). 
Note that in the limit $m \rightarrow
1$ and $\beta = 2$, this solution goes over to the one-soliton solution
\be\label{E99}
u(x,t) = 2\alpha^2 \tanh^2 (\alpha (x-vt))~, ~\alpha = \sqrt{\frac{1-v^2}{8}}~,
\ee  
Now consider the linear superposition
\be\label{1}
u(x,t) = -2 \alpha^2 \sum_{i=1}^{p} d_i^2 (\alpha (x-vt)) + \beta \alpha^2~,
~~(p=1,2,3...)~.
\ee
It is easy to check that this is an exact solution to the Boussinesq 
eq. (\ref{E97}) provided
\be\label{2}
2\big [4-2m -3\beta-6A_1(p,m) \big ]\alpha^2 = v^2 -1~. 
\ee
Here use has been made of the identity
(\ref{E6}) with $A_1 (p,m)$ given by eqs. (\ref{E7}) and (\ref{E9}).
It may be noted that $A_1 (p,m) \le 0$. 

In this article we have shown that in view of the remarkable identities 
satisfied by Jacobi elliptic functions, a kind of linear superposition 
principle works for several nonlinear equations, some of which are associated with integrable systems, the others not. It would indeed be worthwhile to
obtain such solutions for other nonlinear systems where elliptic functions play a role in
the space of exact solutions. A question which comes to mind is how the solutions obtained in this paper are related to previously known solutions. At first sight, it would appear that our procedure has given new solutions, but a closer investigation reveals that our solutions are expressible in terms of previous solutions via a non-trivial generalization of Landen's formulas which connect Jacobi elliptic functions with two different modulus parameters \cite{landen}. The reader is referred to Ref. \cite{landen} for more details.

One of us (A.K.) thanks the Department of Physics at the University of Illinois at Chicago for hospitality. We
acknowledge grant support from the U.S. Department of Energy. 

\end{document}